\input harvmac
\input epsf
\def\journal#1&#2(#3){\unskip, \sl #1\ \bf #2 \rm(19#3) }
\def\andjournal#1&#2(#3){\sl #1~\bf #2 \rm (19#3) }

\def\frac#1#2{{#1\over#2}}

\def\inbar{\,\vrule height1.5ex width.4pt depth0pt}
\def\IC{\relax\hbox{$\inbar\kern-.3em{\rm C}$}}
\def\IR{\relax{\rm I\kern-.18em R}}
\def\IP{\relax{\rm I\kern-.18em P}}
\def\IZ{\relax{\rm I\kern-.18em * 0<
* 1 
* 2 
* 3 
* 4 
* 5 
* 6 
* 7 
* 8 
* 9 
*10 
*11 
*12 
*13 
*14 
*15 
Z}}

%
%

%
\catcode`\@=11
\def\slash#1{\mathord{\mathpalette\c@ncel{#1}}}
\overfullrule=0pt

\def\underrel#1\over#2{\mathrel{\mathop{\kern\z@#1}\limits_{#2}}}

\catcode`\@=12


%


\rightline{RI-08-01}
\Title{
\rightline{hep-th/0108142}}
{\vbox{\centerline{D-Branes on a gauged WZW model}}}
\medskip
\centerline{Shmuel Elitzur and Gor Sarkissian}
\bigskip
\smallskip
\centerline{Racah Institute of Physics, The Hebrew University}
\centerline{Jerusalem 91904, Israel}
\smallskip

\bigskip\bigskip\bigskip
\noindent
The algebraic classification of Cardy for boundary states
 on a $G/H$ coset CFT of a compact group G, is geometrically
 realized on the corresponding manifold resulting from
 gauging the WZW model. The branes consist of H orbits 
of quantized G conjugacy classes shifted
 by quantized H conjugacy classes.

\vfill
\Date{8/01}
\newsec{Introduction}

\lref\klsev{C. Klimcik and P. Severa, Nucl. Phys. B488 (1997) 653.}%
\lref\alschom{A. Alekseev and V. Schomerus, hep-th/9812193.}%
\lref\gaw{K. Gawedzki, hep-th/9904145.}%
\lref\stanold{S. Stanciu, hep-th/9901122, JHEP {\bf 9909} (1999) 028.}%
\lref\bfs{L. Birke, J. Fuchs and C. Schweigert, hep-th/9905038.}%
\lref\gcpleb{H. Garcia-Compean and J.F. Plebanski, hep-th/9907183.}%
\lref\arsold{A. Alekseev, A. Recknagel and V. Schomerus, hep-th/9908040,
JHEP {\bf 9909} (1999) 023.}%
\lref\fffs{G. Felder, J. Frohlich, J. Fuchs and C. Schweigert,
hep-th/9909030.}%
\lref\fofs{J.M. Figueroa-O'Farrill and S. Stanciu, hep-th/0001199.}%
\lref\bdsnew{C. Bachas, M. Douglas and C. Schweigert, hep-th/0003037.}%
\lref\arsnew{A. Alekseev, A. Recknagel and V. Schomerus, hep-th/0003187.}%
\lref\fredshom{S.\ Fredenhagen and V.\ Schomerus, hep-th/0012164.}%
\lref\malmnat{J.\ Maldacena, G.\ Moore and N.\ Seiberg, hep-th/0108100.}%
\lref\cardy{J.L. Cardy, Nucl. Phys. B324 (1989) 581.}%
\lref\brs{K. Bardakci, E. Rabinovici and B. Saring, 
Nucl. Phys. B299 (1988) 151.}
 \lref\ngaw{K. Gawedzki, hep-th/0108044.}%
\lref\dif{P. Di Francesco, P. Mathieu and D. Senechal, Conformal Field Theory, 1997, Springer-Verlag, New York.}%
\lref\malmosei{J. Maldacena, G. Moore and N. Seiberg, hep-th/0105038}%
On a background of conformal field theory branes can be
 described by boundary states. Demanding preservation
 of maximal symmetry, these boundary states were 
classified by Cardy \cardy\  for a rational CFT.
 In that case a boundary state, that is, a type of a D-brane,
 was found to exist in correspondence to each primary field
 of the chiral algebra of the given CFT. Geometrically 
D-branes are embedded as subsets of target space.
 Whenever a CFT target space possesses a geometrical
 interpretation the algebraically constructed Cardy boundary
 states should be realized as such subsets. In the case
 of a CFT which is a compact group manifold such a realization
 exists \alschom\ , \refs{\klsev,\stanold,\gaw,\bfs,\gcpleb,
\arsold,\fffs,\fofs,\bdsnew,\arsnew,\fredshom,\malmnat}.
\lref\gawkup{K.\ Gawedzki and A.\ Kupiainen, Phys. Lett.
{\bf B215} (1988) 119}
\lref\gawkupc{K.\ Gawedzki and A.\ Kupiainen, Nucl. Phys.
{\bf B320} (1989) 625}

  D-branes can sit on a finite set of allowed conjugacy classes
 on the group manifold which are in a natural correspondence
 with the representations of the affine group generated from 
 the primary fields present in the model.

In this work we study a geometrical realization of D branes for
 a more general class of conformal field theories, the coset
 models $G/H$ where $H$ is a subgroup of a compact group $G$. Algebraically
 this is a rational CFT whose allowed boundary states obey
 the Cardy's classification. A geometrical description of
 such models is provided when considered as gauged WZW models
 integrating out the gauge fields \brs, \gawkup, \gawkupc. 

  Following the method applied
 to CFT on a group manifold we construct a  WZW action for
 a world sheet with boundary moving on the group manifold
 $G$ with the subgroup $H$ gauged away. The boundary is
 constrained into some subsets of $G$ dictated by the preserved
 symmetry. The consistency of this action puts some quantization
 conditions on the allowed positions of the branes, establishing
 their correspondence to the primary fields of the coset CFT.
The dimensionalities of the resulting
generic branes are determined.

We limit ourself to maximally symmetric branes. The more general case 
of branes with a smaller symmetry was studied for the parafermionic example
in \malmosei.

We deal with non-abelian $G$ and $H$. The case of $U(1)$
gauge group was discussed extensively recently, in particular
 with respect to the parafermionic example  \malmosei.   

In sec. $2$ the analysis of branes on a group manifold is reviewed.
 Sec. $3$ deals with the  $G/H$ model.
 In sec. $4$ the case of $G$ and $H$ with a common center is discussed.
In an appendix, some formulae from the main text are motivated.

 After completing this work we found the paper \ngaw\ by K. Gawedzki, whose
  sec. 2  contains results similar to ours.

\newsec{Branes on a WZW model} 
\lref\polwie{A.\ M.\ Polyakov and P.\ B.\ Wiegmann,  Phys. Lett. {\bf 131B}
 (1983) 121.}%
\lref\polwieg{A.\ M.\ Polyakov and P.\ B.\ Wiegmann, Phys. Lett. {\bf 141B}
 (1984) 223.}%
As a preparation to
the case of cosets we review here the case of a group manifold. We follow
the discussion of \alschom, where the SU(2) case is dealt with and the
generalization to any compact group in \gaw. We check explicitely the
symmetries of the resulting action.

The action of a WZW model for a world sheet $\Sigma $ without boundary is

\eqn\wzw{ S =S^{\rm kin} + S^{\rm WZ} = 
{ k_{G}\over{4 \pi}}\left[\int_{\Sigma}d^2z L^{\rm kin} 
+ \int_B \omega^{\rm WZ}\right]}

where  $L^{\rm kin} = {\rm Tr}  (\partial_z g \partial_{\bar z } g^{-1} )$ ,
 $\omega^{\rm WZ}={ 1\over 3} {\rm Tr} (g^{-1} dg)^3$, and $B$ is a three-manifold
 bounded by $\Sigma$. Here $g(z,\bar z)$ is the embedding of
 $\Sigma$ into the compact group manifold $G$ and the integer $k_{G}$ is the
  level of the model. This action is invariant
 under the $\hat {G} \times \hat {G}$ symmetry, $\hat {G}$ being the
 loop group of $G$, 

\eqn\sym{g(z,\bar{z})\rightarrow  h_L(z) g(z,\bar{z}) h_R(\bar {z})}

This invariance can be seen by substituting
 $ h_L(z) g(z,\bar{z}) h_R(\bar {z})$ for $g$ in \wzw\
using the Polyakov - Wiegmann identities \polwie\ , \polwieg:

\eqn\pwk{ L^{\rm kin}(g h)=  L^{\rm kin}(g) + L^{\rm kin} (h)
 -( {\rm Tr} (g^{-1}\partial_z g \partial_{\bar z} h h^{-1}+
 {\rm Tr} (g^{-1} \partial_{\bar z} g\partial_z  h h^{-1}))}

\eqn\pwwz{\omega^{\rm WZ}(g h) = \omega^{\rm WZ}(g) + \omega^{\rm WZ}(h)
 - d({\rm Tr} (g^{-1} dg  dh h^{-1}))}

together with the fact that $\partial_z h_R= \partial_{\bar z}h_L = 0$.
  
 When $\Sigma$ has boundaries, left moving and right moving waves
 are mixed by them. The two independent symmetries in \sym\ cannot
 be present. Still one can put boundary conditions which preserve
 the symmetry \sym\ in the bulk with $h_L$ and $h_R^{-1}$ tending
 on the boundary to a common limit. The symmetry on the boundary is then
\eqn\bsym{g\rightarrow h(\tau)g h(\tau)^{-1}}
where $\tau$ parameterizes the boundary.
 The boundary conditions on $g$ should respect the symmetry \bsym\ .
If a group element $f$ is allowed on the boundary, the symmetry implies 
that $kfk^{-1}$ should also be allowed, for any $k\in G$. Thus the boundary
should be allowed to take values in the whole conjugacy class
 $C^{G}_{f}$ defined by 
 \eqn\concl{C^{G}_{f}=\{g\in G | g=kfk^{-1}\}}

Suppose then that the boundary of $\Sigma$ is constrained to map into the 
conjugacy class $C^{G}_{f}$ ,
\eqn\conj{g(\tau)= k(\tau)f k^{-1}(\tau)}
for some fixed group element $f$.
As it stands the action \wzw\ is not well defined for a world sheet with 
boundary. There is no region $B$ bounded by $\Sigma$ when  $\Sigma$ itself
has boundary. To fix that, for a world sheet with a single hole, 
one extends the mapping from $\Sigma$ to $G$ to
a surface without boundary $\Sigma \cup D$ where $D$ is an auxiliary disc 
which closes the hole in $\Sigma$ having a common boundary with it \alschom.
  The disc 
$D$ is mapped into the same conjugacy class \conj\ 
 allowed for its boundary. The 
region $B$ in the action \wzw\  is then taken to be bounded by $\Sigma \cup D$.
The action should be formulated such that the embedding of the interior of 
 the auxiliary disc $D$ inside the conjugacy class does not matter as 
 long as its boundaries are fixed. We also demand that
the symmetry \bsym\ which in general does affect also the boundaries 
of $D$ will continue to be respected by the action. This is done by 
modifying the
 action into:  

\eqn\moact{ S ={ k_{G}\over{4 \pi}}\left[\int_{\Sigma}d^2z L^{\rm kin} 
+ \int_B \omega^{\rm WZ}-
\int_{D}\omega^{f}(k)\right]}

where $\partial B = \Sigma \cup D$ and  $\omega^{f}(k)$ is the two-form 
defined on the conjugacy class \conj\ for $g=kfk^{-1}$ as \alschom,

\eqn\om{\omega^{f}(k) = {\rm Tr}(k^{-1}dk f k^{-1}dk f^{-1})}

On the class \conj\ , where it is defined, $\omega^{f}$ satisfies

\eqn\dw{d\omega^{f} = \omega^{\rm WZ}}

Under the transformation \sym\ with boundary values \bsym\ , the change
in the $L^{\rm kin}$ term in  \moact\ read from \pwk\ ,  is canceled
 by the corresponding $\Sigma$ integral of the  boundary term
 from the change in the $ \omega^{\rm WZ}$ term, read from \pwwz\ . In the presence
 of a world sheet  boundary there remains the contribution from $D$ to 
the latter change, 

\eqn\delwz{\Delta (S^{\rm kin}+S^{\rm WZ}) = 
{ k\over{4 \pi}} \int_D {\rm Tr}[h^{-1}dh( g h^{-1}dh g^{-1} 
- g^{-1}dg - dg g^{-1})]}
   
The change in the $\omega^{f}$ term in \moact\ can be read from the identity
\eqn\chom{\omega^{f}(hk)-\omega^{f}(k)=
{\rm Tr}[h^{-1}dh(gh^{-1}dhg^{-1}-g^{-1}dg-dgg^{-1})]}
where $g=kfk^{-1}$, which follows from the definition \om\ .
  The change in the $\int_{D}\omega^{f}(k)$ term exactly cancels \delwz\ 
  so that \moact\ is indeed
invariant under \sym\ . 

Eq. \dw\ guarantees that the action is invariant under
continuous deformations of the embedding of the auxiliary disc inside
the conjugacy class. The change in the $\omega^{f}$ integral cancels
 the integral
of the $\omega^{\rm WZ}$ term on the three-volume swept by the disc during
 such a deformation. Since in general the second homotopy of a conjugacy 
class is non trivial, there are different embeddings of a disc in such a 
class which are not continuously connected. The union of two such 
embedded discs is not the boundary of a three volume inside 
the conjugacy class, where \dw\ is valid. The action \moact\ is in fact 
sensitive to such a topological change in the embedding of the auxiliary 
disc. In order that this change will have no physical effect, the induced 
change in the action must be an integral multiple of $2\pi$. This leads
\alschom, \gaw\  
  to a quantization of the conjugacy classes  allowed for as boundary
 conditions. Since $k$ in \conj\ is defined modulo right 
 multiplication by any element commuting with $f$ and the 
 group of such elements for a generic $f$ is isomorphic to $T^{G}$, 
 the Cartan torus of $G$, the conjugacy class \concl\ can be 
 described as $G/T^{G}$. Its second homotopy group is therefore,
\eqn\homot{\Pi^{2}(C^{G}_{f})= \Pi^{1}(T^{G})}
  If $r$ is the rank of $G$, a topologically non trivial 
  embedding of $S^{2}$ in $C^{G}_{f}$ is characterized by an $r$ 
  dimensional vector in the coroot lattice of $G$.  Namely,
   if one embedding $D$ of
   the disc into $C^{G}_{f}$ is given by $kfk^{-1}$ and another 
   embedding $D'$ sends it into $k'fk'^{-1}$, then on the 
   topologically circular boundary the two embeddings should coincide. 
   This implies
\eqn\wind{k(\tau)k'(\tau)^{-1}=t(\tau)}
where $t(\tau)$ is an element of the subgroup isomorphic to $T^{G}$
 which commutes with $f$. Eq. \wind\ determines a mapping from 
 the circular boundary of a given hole in the world sheet into 
 the torus $T^{G}$. Since $T^{G}$ is $R^{r}$ modulo $2\pi$ times
 the coroot lattice,
 every such  mapping belongs to a topological 
 sector parameterized by a vector in the coroot lattice describing the winding of
 this circle on the torus $T^G$. 
 This lattice vector determines, by \homot\ ,
  the element of $\Pi^{2}(C^{G}_{f})$ 
 corresponding to the union of $D$ and $D'$.
 
  Let the element f in \conj\   
 chosen in the Cartan torus
be of the form $f= e^{i \theta \cdot \lambda }$ where $\lambda$ are Cartan 
generators. The change in the action resulting from 
 a topological change in the embedding of the disc which is 
 characterized by a coroot lattice vector $s$, 
 is given by \gaw 
\eqn\topc{\Delta S=  k_{G}(\theta \cdot s)}
where the length of long roots is normalized to $2$ . 
Consistency of the model then implies the condition
\eqn\quan{\theta\cdot \alpha \in 2\pi Z/{k_G}}
for any coroot $\alpha$. In this normalization the weight lattice is the set of
points in $R^r$ whose scalar product with any coroot takes integral values.
 Eq. \quan\ implies then, that $\theta$ should be $2\pi/k_{G}$ 
times a vector in the weight lattice. As a point in $T^G$,  $\theta$ is defined
modulo $2\pi$ times the coroot lattice. The allowed conjugacy classes correspond
then to points in the weight lattice divided by $k_G$, modulo the coroot lattice. 
This is also the characterization of the integrable representations of $\hat G$,
the affine $G$ algebra at level $k_G$, which correspond to the primary fields
of the WZW model.
 It is in
accordance with the algebraic analysis of Cardy \cardy\ where
 a correspondence is established between 
primary operators in a rational CFT and boundary states. The above discussion
is a geometrical realization on the group manifold of this correspondence.

It is of course equally consistent and symmetric to replace the set
of conjugacy classes \conj\ with quantization condition \quan\ , by the 
same classes shifted by a fixed group element $m$. The boundary conditions
\conj\ are then modified into  

\eqn\sconj{g(\tau)= k(\tau)f k^{-1}(\tau)m}
where f satisfies \quan\ and m is an arbitrary fixed group element. 
This amounts to
constrain the boundary values of $g$ into the shifted conjugacy class
 $C^{G}_{f}m$. This set
of boundary conditions preserves a boundary symmetry different
from  \bsym\ . The shifted symmetry is 

\eqn\sbsym{g\rightarrow h(\tau)g m^{-1} h(\tau)^{-1} m }

If we insist on the boundary symmetry \bsym\ , the allowed boundary states 
are the
set \conj\ with the condition \quan.

\newsec{Branes on a Coset}
\lref\kpsy{D.\ Karabali, Q-H Park, H.\ Schnitzer and Z.\ Yang, Phys. Lett. 
{\bf B216} (1989) 307.}%
\lref\ks{D.\ Karabali and H.\ Schnitzer, Nucl. Phys. {\bf B329} (1990) 649.}%

Let $G$ be a compact, simply connected, non-abelian group. 
The $G/H$ coset CFT, where $H$ is a subgroup of $G$, can be described 
in terms of a gauged WZ action \brs\ ,
where the symmetry 

\eqn\gauge{g\rightarrow hgh^{-1}}

$g\in G$ , $h\in H$ is gauged away. An $H$ Lie algebra valued
 world sheet vector field $A$  is added to the system, and the $G/H$ 
 action  on a world sheet without boundary becomes,

\eqn\gact{\eqalign{& S^{G/H} = S^{\rm kin} + S^{\rm WZ} +S^{\rm gauge}\cr & = 
{ k_{G}\over{4 \pi}}\left[\int_{\Sigma}d^2z L^{\rm kin} 
+ \int_B \omega^{\rm WZ}\right] \cr &+  
{k_{G}\over{2 \pi}}\int_{\Sigma} d^{2}z {\rm Tr}( A_{\bar z}
\partial_{z} g g^{-1} - A_z g^{-1} \partial_{\bar z} g +A_{\bar z}g A_z g^{-1}
-A_z A_{\bar z})}}

Introduce $H$ group element valued world sheet fields $U$ and $\tilde U$ as
\kpsy, \ks,

\eqn\subs{\eqalign{&A_z = \partial_z U U^{-1}\cr &A_{\bar z} =
 \partial_{\bar z} \tilde U {\tilde U}^{-1}}}

Denote the action \wzw\ by $S^{G}(g)$ . Then the coset action \gact\ becomes,

\eqn\uact{S^{G/H}= S^{G}(U^{-1}g \tilde {U})- S^{H} (U^{-1} \tilde {U})}
as can be checked using Polyakov Wiegmann identities \pwk\ , \pwwz\ and 
the definitions \subs\ . The level $k_{H}$ of the $S^{H}$ term in \uact\ , 
is related to $k_{G}$ through the embedding index of $H$ in $G$ \dif .

The model has then the following symmetries. First of all one should identify 
configurations related by the local gauge transformation,
\eqn\gautr{\eqalign{
g(z,\bar {z})&\rightarrow h(z,\bar {z})g(z,\bar {z})h^{-1}(z,\bar {z})\cr
                          U(z,\bar {z})&\rightarrow h(z,\bar {z})U(z,\bar {z})
			  \cr
                         \tilde {U}(z,\bar {z})&
\rightarrow h(z,\bar {z})\tilde {U}(z,\bar {z})}}
with $h(z,\bar {z})\in H$.
In addition, by \sym\ , there are further global  
symmetries
\eqn\cosym{\eqalign{
U(z,\bar {z})&\rightarrow U(z,\bar {z}) h_L^{-1}(z)\cr
           \tilde {U}(z,\bar {z})&\rightarrow 
	   \tilde {U}(z,\bar {z}) h_R(\bar z)\cr
           g(z,\bar {z})&\rightarrow g(z,\bar {z})}}
where $h_L, h_R \in H$.
 The action is also invariant under
\eqn\gsym{\eqalign{ g(z,\bar {z})&\rightarrow
 m_L(z)g(z,\bar {z})m_{R}(\bar z)\cr
U&\rightarrow U\cr \tilde {U}&\rightarrow \tilde{U}}}
with $m_L,m_R \in G$ and $[m_L,H] = [m_R,H] = 0$ .

Suppose now that we gauge the WZW model of the group $G$,
 defined on a world sheet with boundary with
 the boundary conditions \conj\ on $g(z,\bar z)$.  
 On the group manifold, we saw in previous section that 
 boundary conditions corresponding to shifted conjugacy 
 classes \sconj\ did not preserve the same symmetry of the 
 condition \conj\ but rather a different symmetry \sbsym\ . 
 Insisting on the symmetry \sym\ we did not include this kind of conditions. 
 Here on the coset, the symmetry  \gsym\ is  limited
relative to \sym\ , and we note that shifted boundary 
conditions like \sconj\ are still consistent with them 
provided the shift element $m$ in \sconj\ belongs to $H$. 
We will then consider a more general type of 
boundary conditions on $g$ allowing on the boundary
\eqn\sconjh{g(\tau)= k(\tau)f k^{-1}(\tau)l}
with $f,k\in G$ and $l$ an element in $H$.
For a fixed $l$, these boundary conditions are not gauge
invariant. The gauge symmetry \gautr\ forces us to allow,
together with shifting by $l$, to shift by any element in
 its $H$ conjugacy class. We are then lead to the following 
 boundary conditions,
 \eqn\pcon{g(\tau)= k(\tau)f k^{-1}(\tau) p(\tau)lp^{-1}(\tau)}
 with $p(\tau), l\in H$. In other words, on the boundary
  $g$ is constrained to a product $ C^{G}_{f} C^{H}_{l}$ 
  of a $G$ conjugacy class with an $H$ class.

As in previous section the presence of the WZ term
necessitates an auxiliary disc $D$ for each hole in the world sheet,
to define the boundary of the three-volume integral in the action.
 To proceed, we will continue
from the boundary into the disc $D$ the field $g$ subject to conditions 
\pcon. 
 Here again an additional two-form
has to be added on the disc to give meaning to the action. 
For the product of classes boundary conditions \pcon\ the two-form 
introduced in \om\ is not appropriate. 
 To complete the action define another two-form as follows.
  Let $c_{1} = k f k^{-1}$, $f\in G$, and $c_{2} = p l p^{-1}$,
   $l\in H$. Define
\eqn\gom{\Omega^{(f,l)}(k,p) =  \omega^f (k)+
 {\rm Tr}(dc_{2} c_{2}^{-1} c_{1}^{-1} dc_{1}) + 
\omega^l (p)}
where $\omega^f (k)$ and $\omega^l (p)$ are two-forms defined
 as in \om. Considerations leading to this particular 
form are discussed in the appendix.
 We will then add to the action \gact\ the term 
\eqn\disct{-{k_{G}\over{4\pi}}\int_{D}\Omega^{(f,l)} (k,p)}
To check the consistency of the action with the term \disct\ , 
notice first that for $g=c_{1}c_{2}$, 
the WZ term has the form
\eqn\wzcc{\eqalign{\omega^{\rm WZ}={1\over{3}}{\rm Tr} (c_{1}^{-1}dc_{1})^{3}+
&{1\over{3}}{\rm Tr} (c_{2}^{-1}dc_{2})^{3}+
 {\rm Tr}(c_{1}^{-1}dc_{1})^{2}dc_{2}c_{2}^{-1} 
+{\rm Tr}(c_{1}^{-1}dc_{1})(dc_{2}c_{2}^{-1})^{2}}}
Using \dw\ one gets,
\eqn\dwcc{d\Omega^{(f,l)}= \omega^{\rm WZ}} 
This guarantees the invariance of the action \gact\
with the additional term \disct\ under a continuous
 deformation of the embedding of the disc $D$ into 
 $G$ subject to conditions \pcon. 

As to the symmetry \gsym\ with $m_{L}=m_{R}^{-1}=m$ on the boundary, 
it acts on the term \disct\ 
 taking $c_1\rightarrow mc_1 m^{-1}$ for $[m,H]=0$. 
The change in $\Omega^{(f,l)}$ is 
given by
\eqn\chcaom{\Delta \Omega^{(f,l)} =
 {\rm Tr}[(m^{-1}dm)(c_{1}m^{-1}dm c_{1}^{-1} 
- dc_{1}c_{1}^{-1}-c_{1}^{-1}dc_{1}-dc_{2}c_{2}^{-1}
-c_{1}dc_{2}c_{2}^{-1}c_{1}^{-1})]}
This exactly cancels the variation of the bulk terms read 
from \delwz\ when $m$ is substituted there for $h$, $c_{1}c_{2}$
 for g and the commutation of $m$ with $c_{2}$ taken into account.

 As in previous section, the embedding of the disc $D$
  into $ C^{G}_{f} C^{H}_{l}$ involves a topological
 choice. Holding  $plp^{-1}$ in \pcon\ fixed on the disc while performing a 
 topological change corresponding  to a $G$ coroot
 lattice vector $s_{G}$ in the definition on the interior of $D$,
 of the factor $kfk^{-1}$ ,
  will induce in $S$  in \gact\ the same change as that of 
  previous section
 \eqn\topcg{\Delta_{G} S = k_G(\theta_G \cdot s_G)}
 where $f=e^{i \theta_G\cdot \lambda_G}$. 
 The consistency of the action requires
 then the same quantization 
 condition \quan\ on the $G$ conjugacy class 
 \eqn\quang{\theta_{G}\cdot \alpha_G \in 2\pi Z/{k_G}}
 Similarly, a topological change corresponding to an $H$ coroot
 vector $s_{H}$ in the continuation to $D$ of the factor $plp^{-1}$ in 
 \pcon\ with the $kfk^{-1}$ held fixed, will also change $S$.
 For $l=e^{i\theta_{H}\lambda_{H}}$ this change will be
 \eqn\topch{\Delta_{H} S = k_H(\theta_H \cdot s_H)}
 The consistency of the action \uact\ then also constrains the 
 $H$ conjugacy class factor by 
 \eqn\quanh{\theta_{H}\cdot \alpha_H \in 2\pi Z/{k_H}.}
 
 There is a problem in the two-form $\Omega^{(f,l)}(k,p)$ 
 introduced into the action in eq. \disct. This form depends explicitly 
 on the $G$ and $H$ classes factors of the group element $g$ 
 on the boundary and in the disc. In general, however, the 
 factorization of an element $g \in C^{G}_{f}C^{H}_{l}$ into a product 
 of the form $g=kfk^{-1} plp^{-1}$ is not unique. Let $c_1 = kfk^{-1}$
 and $c_2 = plp^{-1}$ . Varying $k$ infinitesimally by 
 \eqn\varone{\delta k = q_{G} k}
 and $p$ by
  \eqn\vartwo{\delta p = q_{H} p}
   where $q_{G}$ and $q_{H}$ 
 belong to $G$ and $H$ Lie algebras respectively, $c_1$ is changed by
 $\delta c_1 = [q_G,c_1]$ and $c_2$ varies  by
   $\delta c_2 = [q_H,c_2]$ . If the Lie algebra 
   elements $q_{G}$ and $q_{H}$  are chosen such that they satisfy
   \eqn\amb{q_{G} - c_1^{-1} q_{G} c_1 = q_{H} - c_2 q_{H} c_2^{-1}}
   then the product $g = c_1 c_2 $ is unchanged under the above variation.
   The left hand side of eq. \amb\  is a linear operator in the
 Lie algebra of $G$ acting on $q_{G}$. This operator has a null
   space of dimension $r_G$ , the rank of $G$, namely the $r_G$ independent
   elements commuting with $c_1$ . Choosing $c_1$ 
   in the Cartan
   torus of $G$ this null space is the Cartan subalgebra of $G$.
   For a $q_H$ in the $H$ Lie algebra,
   which does not commute with $c_2$, there will be a $q_G$ solving \amb\
   provided that $q_{H} - c_2 q_{H} c_2^{-1}$ has 
   no component in this Cartan algebra. The dimension
   of the set of pairs $(q_G,q_H)$ solving \amb\ , is the dimension of 
   the set of elements $q_H$  not counting their components
    along the Cartan algebra
   commuting with $c_2$, such that  $q_{H} - c_2 q_{H} c_2^{-1}$
   has no components in the algebra commuting with
   $c_1$. This is the dimension
   of different factorizations of a given $g = c_1 c_2 $
   in  $C^{G}_{f}C^{H}_{l}$  as a product of 
   a pair of elements from both classes.
    For generic $c_1$ and $c_2$, for which the two Cartan algebras
commuting with $c_1$ and $c_2$ are disjoint,
   this dimension is $d_H-2r_H$. At a 
   generic  $g = c_1 c_2 $ point in the region $C^{G}_{f}C^{H}_{l}$ its
   dimension is then given by
   \eqn\dpr{d(C^G_f C^H_l) = d_G -r_G + d_H-r_H  - (d_H-2r_H)
    = d_G - r_G + r_H}
    Non generic points for which the algebra commuting with $c_1$ and 
    that commuting with $c_2$ have a common subspace, form lower dimensional
    boundary in $G$ of the region $C^{G}_{f}C^{H}_{l}$.
    
    Since fixing $g \in C^{G}_{f}C^{H}_{l}$ does not determine $c_1$ 
    and $c_2$ the two-form $\Omega$ in the action \disct\ seems to require 
    more data than just $g$. In fact one can check the behavior of 
    $\Omega^{(f,l)}(k,p)$ under the variation \varone\ and \vartwo\
    and find that it is not invariant even when $q_G$ and $q_H$ do
    satisfy \amb\ . Explicitly, in this case we find
    \eqn\varom{\delta \Omega^{(f,l)}(k,p) = -2 d [{\rm Tr}( q_G dc_1 c_{1}^{-1} 
    + q_H c_{2}^{-1} d c_2)]}
    It seems then that for the action \disct\ to make sense, one should 
    allow for extra degrees of freedom on the boundary, fixing the
    factorization of $g$ there as a product of $G$ and $H$ conjugacy
    classes. Namely, the Physics does depend on $k$ and $p$ in \pcon,
    not just on the value of $g$. Note though, that there is no dependence 
    on the factorization chosen inside the disc, once it is fixed 
    on the boundary. Since the variation of $\Omega^{(f,l)}$ in \varom\ has
    the form of a  derivative of a local one form, as could be expected 
    from \dwcc, the variation vanishes for $q_G$ and $q_H$ zero
    on the boundary\foot{See \ngaw\ for a  discussion of this issue in 
the framework of the canonical formalism.}.

The allowed D branes for the $G/H$ model correspond then to a
pair of quantized conjugacy classes of the two groups. Since each
such a class corresponds to an integrable representation, we get a
characterization of these branes by pairs of a $G$ primary field and an
$H$ primary field. This is again in accordance with Cardy's analysis, the 
primary fields of the $G/H$ CFT, generically correspond to such pairs.

    A geometric picture for the $G/H$ coset model emerges once the gauge is 
    fixed and the gauge fields are integrated over \brs\ . 
    The points of the resulting manifold can be identified as 
    the orbits in $G$ of the gauge transformation \gauge\ .
    The boundary conditions \pcon\ put the branes on $G$ conjugacy classes
    in the $G$ group manifold, quantized according to \quang\ ,
     shifted, in the sense of \sconj\ ,
    by  $H$ conjugacy classes, which are also quantized according to \quanh\ .
    All the points on the orbit of a gauge transformation \gauge\ are
    identified. For generic $f\in G$ and $l \in H$ 
    the dimension of the product $C^G_f C^H_l$ at a generic point is
    given in eq. \dpr\ .
    Identifying the $H$ gauge
    orbits generically reduces the dimension by $d_H$. The generic
    dimension of the geometrically realized $D$ branes  is then, for
    non abelian $G$ and $H$,
    \eqn\dbr{{\rm dim}({\rm brane})=d_G-d_H -r_G +r_H.}

\newsec{The case of a common center}
\lref\dgepiden{D.\ Gepner, Phys. Lett. {\bf B222}, (1989), 207.}
\lref\dgepscal{D.\ Gepner, Nucl. Phys. {\bf B322}, (1989), 65.}  

When $H$ contains some subgroup $C$ of the center of $G$, 
the above discussion gets modified in two ways. 
First, for $z \in C$  the region $C^{G}_{f}C^{H}_{l}$ is identical to
the region $C^{G}_{zf}C^{H}_{z^{-1}l}$. 
The brane corresponding to the pair $(f,l)$ of conjugacy classes
is then identical to the brane corresponding to the pair
$(z f, z^{-1} l)$. This is the geometrical origin of the phenomena 
known in the context of coset CFT without boundary
 as "field identification" \dgepiden,\dgepscal,\dif\ .
It is again consistent with Cardy's identification of boundary states with
 primary fields.

Since the  gauge transformation takes $g$ into $hgh^{-1}$, it does not
distinguish between the transformations $h$ and $zh$  for any 
$z \in C$. We can then think of the gauge group as $H/C$. Recall 
that the element $k$  in \pcon\ is defined modulo right
multiplication by $G$ group elements from the torus $T^{G}_{f}$ 
commuting with $f$. Similarly $p$ in that 
equation can be multiplied from the right by
any element of $T^{H}_{l}$. Let the boundary of the hole be 
parameterized by $0\le \tau \le 2\pi$. We have seen 
in \wind\ that upon replacing the boundary value $k(\tau)$ by
\eqn\redk{k'(\tau) = k(\tau)t(\tau)}
 with
\eqn\twr{t(\tau)= e^{{i\over{2\pi}}\tau (s \cdot \lambda)}}
 $s$ being a
coroot lattice vector and $\lambda$ a vector of generators commuting 
with $f$, continuing $k'$ rather than $k$ into the disc, 
 the change \topcg\ is induced in the action. This gave rise to 
the quantization condition \quang.
 A similar independent change in $p(\tau)$
induces the change \topch\ leading to the condition \quanh.
 Recall also that a gauge transformation $h \in H$ multiplies both $k$ 
and $p$ by $h$ from the left. Let $z \in C$ be represented as
\eqn\cent{z=e^{i(w \cdot \lambda)}.}
Notice that $w$ is a common weight vector of $G$ and $H$.
Consider an $H/C$ gauge transformation $h(z,\bar z) \in H$,
 which satisfies on the boundary of the hole 
\eqn\tga{ h(0)=z^{-1}h(2\pi)}
Let this transformation act on a configuration with a given continuous
 choice of $k$ and $p$ on the boundary and inside the disc.
 On the world sheet $\Sigma$ the action density, being gauge invariant,
does not change. The representation \pcon\ of $g$ on the boundary is
changed, the transformed $k$ and $p$ satisfy
\eqn\trkp{\eqalign{&k(0)=z^{-1}k(2\pi)\cr&p(0)=z^{-1}p(2\pi)}}
 In this form $k$ and $p$ are discontiuous in $H$. They are
continuous in $H/C$, but the paths $k(\tau)$ and $p(\tau)$ of
\trkp\ are non contractible in $H/C$ and cannot be continued into
the interior of the disc to be substituted in the action \disct.
To define the action we must, before continuing into the disc, 
to redefine $k$ and $p$ according to \redk, multiplying them
 from the right by an appropriate Cartan element, changing
$k$ into $k'$ and $p$ into $p'$ defined as
\eqn\twkp{\eqalign{&k'(\tau)=k'(\tau) e^{{i\over{2\pi}}\tau (w \cdot \lambda)}
\cr&p'(\tau)=p'(\tau) e^{{i\over{2\pi}}\tau (w \cdot \lambda)}}}
The redefined $k'$ and $p'$ are contractible and can be continued into the
disc. The redefinition \twkp, like \redk, induces a change in the disc 
term of the action, according to \topcg\ and \topch. Notice
that, unlike \twr, \twkp\ contains a weight vector rather than a
root vector, and that this twist is done together on $k$ and on $p$.
Equations \topcg\ and \topch\ give then for the
 change of the action  induced by \twkp\ 
\eqn\topcen{\Delta S= (k_{G} \theta_{G} + k_{H} \theta_{H}) \cdot w}
where $f=e^{i(\theta_{G} \cdot \lambda)}$ and
$l=e^{i(\theta_{H} \cdot \lambda)}$.
Invariance under the gauge transformation \tga\ requires this change to 
be a multiple of $2\pi$ leading to a further condition,
 a correlation between $G$ and $H$ conjugacy classes,
\eqn\quacen{(k_{G} \theta_{G} + k_{H} \theta_{H}) \cdot w \in 2\pi Z}
for every common weight of $G$ and $H$.
This is again in accordance with Cardy's correspondence of boundary
states with primary fields of the CFT without boundary. The condition
\quacen\ for coset CFT is known as the selection rule \dgepiden,\dgepscal,
\dif\ , demanding
the same behavior of members of the pair of $G$ and $H$ representations
under the common center.

\vfill\eject

\bigskip
\noindent{\bf Acknowledgments:}
We thank M. Berkooz, A. Giveon, D. Kazhdan, D. Kutasov, E. Rabinovici,
N. Seiberg and A. Schwarz  for useful
discussions. This work is supported in part by the Israel Academy
of Sciences and Humanities -- Centers of Excellence Program.

\appendix{A}{}

Here we present an alternative reasoning bringing to the boundary condition
\pcon\ and the two-form \gom.

Let us study the action \gact\ in the presence of boundary with
the boundary condition \conj\ modified with boundary
 term $\int_D\omega^{f}(k)$:
\eqn\gmact{\eqalign{
&S={k_G\over{4\pi}}\left[\int_{\Sigma}d^2z L^{{\rm kin}}
+\int_B\omega^{{\rm wzw}}
-\int_D\omega^{f}(k)\right]\cr
& + {k2\over{\pi}}\int_{\Sigma}d^2z{\rm Tr}(A_{\bar{z}}\partial_{z}gg^{-1}
-A_{z}g^{-1}\partial_{\bar{z}}g+A_{\bar{z}}gA_{z}g^{-1}-A_{z}A_{\bar{z}})\cr}}
where $g|_D=c_1=kfk^{-1}$.
Using again the parameterization \subs, by straitforward computation we can
show that 
\eqn\muact{\eqalign{
&S=S^{G}(U^{-1}g\tilde{U})-S^{H}(U^{-1}\tilde{U})\cr
&-\int_D\left[\omega^{f}(U^{-1}k)+
{\rm Tr}(d(U^{-1}\tilde{U})(U^{-1}\tilde{U})^{-1}
(U^{-1}c_1U)^{-1}d(U^{-1}c_1U))\right]\cr}}
Till now we did not specify boundary conditions  for the gauge field.
 Denote $U^{-1}\tilde{U}=h$.
 We demand that at the boundary 
\eqn\ubc{h=U^{-1}\tilde{U}|_D=c_2=plp^{-1}}
which is consistent with \gautr\ and
\cosym\ for $h_{L}=h_{R}^{-1}$ .
Adding and subtracting to the action \muact\ the expression
 $\int_D\omega^{l}(p)$
we can write it in the form:
\eqn\muacttrans{\eqalign{
&S = S^{G}(U^{-1}g\tilde{U})-\int_D\left[\omega^{f}(U^{-1}k)+{\rm
Tr}(dc_2c_2^{-1}(U^{-1}c_1U)^{-1}d(U^{-1}c_1U))+\omega^{l}(p)\right]\cr
 &- \left[S^{H}(h)-\int_D\omega^{l}(p)\right]\cr}}
Defining the two-form 
\eqn\gom{
\Omega^{(f,l)}(k,p)=\omega^{f}(k)
+{\rm Tr}\left(dc_2c_2^{-1}c_1^{-1}dc_1\right)+\omega^{l}(p)}
where as before $c_1=kfk^{-1}$ and $c_2=plp^{-1}$, we can
compactly re-write \muacttrans\ as follows:
\eqn\finact{
S=S^{G}(U^{-1}g\tilde{U})-\int_D\Omega^{(f,l)}(U^{-1}k,p)
-\left[S^{H}(h)-\int_D\omega^{l}(p)\right]}
Note that the two-form \gom\ shows the same behavior under 
a shift of its arguments as that of $\omega^f$ in 
\chom, namely
\eqn\trangom{\Omega^{(f,l)}(hk,hp)-\Omega^{(f,l)}(k,p)=
{\rm Tr}[h^{-1}dh(gh^{-1}dhg^{-1}-g^{-1}dg-dgg^{-1})],}
where $g=kfk^{-1}plp^{-1}$.
Finally rearranging terms in \finact\ and
\gmact\
 we get 
\eqn\fgbact{\eqalign{
&\left[S^{G}(g)-\int_D\omega^{f}(k)\right] + 
\left[S^{H}(h)-\int_D\omega^{l}(p)\right]=
S^{G}(U^{-1}g\tilde{U}) -\int_D\Omega^{(f,l)}(U^{-1}k,p)\cr
&-{k_G\over{2\pi}}\int_{\Sigma}d^2z{\rm Tr}(A_{\bar{z}}\partial_{z}gg^{-1}
-A_{z}g^{-1}\partial_{\bar{z}}g+A_{\bar{z}}gA_{z}g^{-1}-A_{z}A_{\bar{z}})\cr}}
We see that the sum of the two actions of the WZW models with boundary 
can be derived after gauging the action given by the first two terms of the
right hand side of \fgbact. Let us study this action in more details.
First of all we see that from the conditions that $g$ and $U^{-1}\tilde{U}$
lie at the boundary in the conjugacy classes $kfk^{-1}$ and $plp^{-1}$
respectively, follows that argument of the action in r.h.s. of \fgbact\
$U^{-1}g\tilde{U}$ lies in the product of these conjugacy classes:
\eqn\prodcl{U^{-1}g\tilde{U}|_D=U^{-1}gUU^{-1}\tilde{U}|_D
=U^{-1}kfk^{-1}Uplp^{-1}=(U^{-1}k)f(U^{-1}k)^{-1}plp^{-1}}
From the fact that the actions in the l.h.s. of \fgbact\ do not change under
infinitesimal variations of the location of the auxiliary disc
 we expect that also the
action on the r.h.s. does not change under them. i.e. as we explained
in section 2 $d\Omega^{f,l}=\omega^{\rm WZ}$ on the product of
 conjugacy classes,
which was explicitly checked in section 3.

\listrefs
\end